# An electronic ratchet is required in nanostructured intermediate band solar cells

Amaury Delamarre, Daniel Suchet, Nicolas Cavassilas, Yoshitaka Okada, Masakazu Sugiyama and Jean-François Guillemoles

*Abstract*— We investigate in this letter the intrinsic properties that have limited the efficiency of nanostructured intermediate band solar cells. Those devices take advantage of intra-band transitions, which occur on narrow energy width, and present low radiative recombination efficiency. We derive the minimum requirements in terms of those two characteristics to achieve efficiencies in excess of the Shockley-Queisser limit, and show that compatible nanostructures are challenging to obtain. Especially, we evidence that currently experimentally considered materials cannot overcome the best single junction cells. In order to solve those issues, we consider devices including an electronic ratchet mechanism. Firstly, such devices are shown to be much less sensitive on the limitations of the nanostructures characteristics, so that requirements for high efficiencies can be met. Secondly, we show that quantum well devices present advantages over their quantum dots counterparts, although they have attracted much less interest so far.

## I. INTRODUCTION

The single junction solar cell efficiency, under one sun illumination, has been shown to be limited by the Shockley-Queisser (SQ) limit, of about 33% [1]. Several strategies have been suggested to overcome this limit, one of them being the concept of Intermediate Band Solar Cell (IBSC) [2], [3], where energy levels are introduced within a semiconductor bandgap, to allow the sequential absorption of sub-bandgap energy photons that would be lost otherwise. Among the different strategies to introduce intermediate levels [4]–[6], a promising one is to take advantage of nanostructures such as Quantum Dots (QDs) or Quantum Wells (QWs).

Nevertheless, to date, no IBSC overcoming the most efficient single junction cells has been demonstrated. Two reasons can be inferred from previous practical realizations. Firstly, using nanostructures, a severe open-circuit voltage ($V_{OC}$) drop is commonly observed when sub-gap states are introduced [5], [6], which denotes strong recombinations. For intrinsic (Auger recombination [7], efficient thermalization in QWs [8] or via multiple states in QDs [9]) or extrinsic (growth issues when going towards large dot densities [4]) reasons. Secondly, low sub-gap currents are observed, due to the monochromatic nature of the inter-sub-band transitions, or to the difficulty of obtaining large dot densities [4]–[6]. Although nanostructured devices have been widely studied since the theoretical description of the IBSC [4]–[6], no systematic study on the interplay of those two parameters on the efficiency has been proposed. We derive in this letter the corresponding minimal requirements in order to overcome the SQ limit, which can be used as guidelines for screening candidate nanostructures. We highlight that currently investigated systems, and especially the widely studied InGaAs QDs in GaAs (e.g. [4]–[6], [10]–[12]), are far from meeting those minimal requirements.

Since nanostructures compatible with high efficiencies appear challenging to fabricate, we consider in addition the possibility to introduce a ratchet mechanism, a concept that can be applied to any energy converter, and in particular to mesoscopic systems [13]. When implemented in an IBSC, a ratchet has been shown to result in an efficiency increase in the radiative limit (from 46.7 to 48.5%) [14]. In addition, it is expected to reduce the harmful effects of absorption reduction [15]. However, its impact in the presence of non-radiative recombination has not been thoroughly investigated [16], despite the fact that the issue of $V_{OC}$ preservation is a common issue of most practical realization [5], [6]. We will show that a ratchet mechanism strongly relaxes the requirements on the absorption and radiative efficiency, to the point that any combination of those two parameters is compatible

A part of this paper is supported by the Research and Development of Ultra-High Efficiency and Low-Cost III-V Compound Semiconductor Solar Cell Modules project under the New Energy and Industrial Technology Development Organization. A part of this paper is supported by a Grant-in-Aid from the Japan Society for the Promotion of Science (JSPS).

Amaury Delamarre, Yoshitaka Okada and Masakazu Sugiyama are with NextPV, LIA CNRS-RCAST/U. Tokyo-U. Bordeaux, Tokyo 153-8904, Japan, and also with the Research Center for Advanced Science and Technology, The University of Tokyo, Tokyo 153-8904, Japan (e-mail: delamarre@hotaka.t.u-tokyo.ac.jp, okada@mbe.rcast.u-tokyo.ac.jp, sugiyama@ee.t.u-tokyo.ac.jp).

Daniel Suchet is with NextPV, LIA CNRS-RCAST/U. Tokyo-U. Bordeaux, Tokyo 153-8904, Japan and also with LPICM, CNRS/Ecole Polytechnique

Universite Paris-Saclay, Palaiseau 91128, France (e-mail: daniel.suchet@polytechnique.org).

Nicolas Cavassilas is with NextPV, LIA CNRS-RCAST/U. Tokyo-U. Bordeaux, Tokyo 153-8904, Japan and also with Aix Marseille Université, CNRS, Université de Toulon, IM2NP – UMR 7334, Marseille, France (e-mail: nicolas.cavassilas@im2np.fr).

Jean-François Guillemoles is with NextPV, LIA CNRS-RCAST/U. Tokyo-U. Bordeaux, Tokyo 153-8904, Japan and also with IPVF, CNRS, UMR 9006, 30 route départementale 128, 91120, Palaiseau, France (e-mail: jf.guillemoles@cnrs.fr).

with efficiencies in excess of the SQ limit. Moreover, this perspective allows us to conclude that QWs present an advantage over QDs for realizing IBSCs, due to their increased ability to form mini-bands that enlarge the sub-gap absorption width. This compensates the increased non-radiative recombination of QWs, when compared to QDs, which is the reason why they have been discarded [17] and much less investigated so far.

## II. THEORETICAL MODEL

The model developed in this paper is similar to those used in previous publications [14], [15], to which non-ideal behaviors, expected from nanostructures, will be introduced. We consider the IBSC represented in Figure 1, in which the CB is split into two distinct bands CB1 and CB2, separated by the ratchet energy $\Delta E$. We assume that the relaxation from CB1 to CB2 is fast, so that both bands share a common quasi-Fermi level. We note $Eg_1$ the bandgap from VB to CB2, $Eg_2$ the bandgap from VB to IB, and $Eg_3$ the bandgap from IB to CB1. We assume $Eg_1 > Eg_2 > Eg_3$. Transitions from CB2 to IB are forbidden. We assume that $Eg_3$ is related to intersubband transitions, so that non-idealities will be introduced relatively to this bandgap. In the following, an IBSC including a ratchet mechanism will be referred to as a RBSC, whereas we will continue to refer to the classical device, without ratchet (i.e. for $\Delta E = 0$ eV), as an IBSC. We note that including a ratchet on the conduction band, valence band or intermediate band is strictly equivalent when calculating the device efficiency limit, which only depends on the three bandgap values [15].

The carrier generation $G_i$ is given by integrating the solar flux multiplied by the absorption of each subgap:

$$G_i = \frac{f}{4\pi^3\hbar^3 c_0^2} \int_0^\infty a_i(E) \frac{E^2}{exp\left(\frac{E}{kT_{sun}}\right) - 1} . dE \qquad (1)$$

$f$ is a geometrical factor, that we choose equal to $6.79*10^{-5}$ for an unconcentrated generation. $a_i(E)$ is the absorption for each bandgap. We assume an ideal absorption for $Eg_1$ and $Eg_2$, i.e. $a_1(E) = 1$ only above $Eg_1$, and $a_2(E) = 1$ strictly between $Eg_2$ and $Eg_1$. In an ideal case, $a_3(E) = 1$ strictly between $Eg_3$ and $Eg_2$. We will study in the following the impact of narrow absorption, in which $a_3(E) = 1$ is limited to the energy range from $Eg_3$ to $Eg_3 + \delta E$.

The radiative recombination $R_i$ for each bandgap $i$ is given by a generalized Planck's law [18]:

$$R_i = \frac{1}{4\pi^2\hbar^3 c_0^2} \int_0^\infty a_i(E) \frac{E^2}{exp\left(\frac{E - \Delta\mu_i}{kT_{cell}}\right) - 1} . dE \qquad (2)$$

The free energy $\Delta\mu_i$ is equal to the quasi-Fermi level splitting in each bandgap. We use the temperatures $T_{sun} = 6000K$ for the sun spectrum and $T_{cell} = 300K$ for the cell. In order to investigate the influence of non-radiative recombinations, we introduce a radiative recombination ratio $r_{rad}$ that defines the global recombination rate for $Eg_3$ following:

$$R_3^* = \frac{1}{r_{rad}} R_3 \qquad (3)$$

We will assume that the recombinations in $Eg_1$ and $Eg_2$ remain radiative. The quasi-Fermi level splittings are related to the voltage V following:

$$qV = \Delta\mu_1 = \Delta\mu_2 + \Delta\mu_3 \qquad (4)$$

The continuity equation in the IB reads:

$$G_2 - R_2 - G_3 + R_3^* = 0 \qquad (5)$$

The extracted current is the difference between the carrier generation and recombination in the conduction band:

$$\frac{J}{q} = G_1 - R_1 + G_3 - R_3^* \qquad (6)$$

The above equations permit to compute IV curves and efficiencies for given sets of bandgaps $Eg_1$, $Eg_2$, $Eg_3$ and the ratchet energy $\Delta E$.

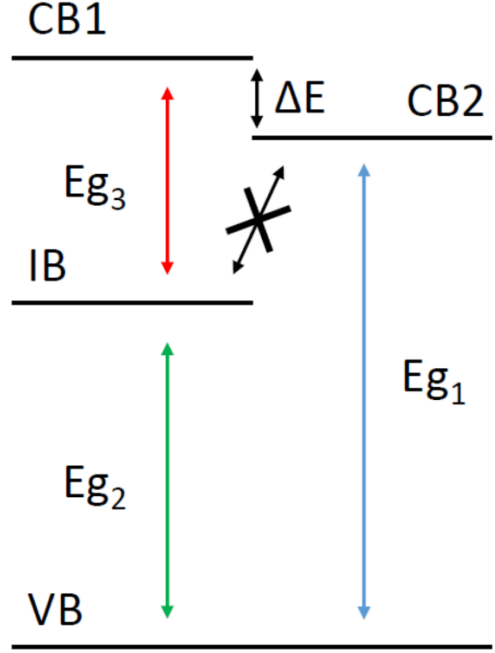

Figure 1 Schematic of an IBSC including a ratchet structure. Compared to a classical IBSC, the conduction band is divided into CB1 and CB2, separated by a ratchet energy ΔE. CB1 is radiatively connected to IB, CB2 is radiatively connected to VB, while no transition from CB2 to IB is allowed. The relaxation structure can be equivalently implemented on the intermediate or the valence band.

## III. RESULTS AND DISCUSSION

The simulation results, with varying absorption width $\delta E$ and radiative efficiencies $r_{rad}$ are presented in Figure 2 and Figure 3. For a first understanding of the results, the efficiencies of ideal devices (RBSC, IBSC and single junction) are compared in Figure 2 as a function of Eg₁. For each value of Eg₁; Eg₂, Eg₃ and ΔE are optimized. As pointed out in [14], the ratchet allows for some improvement compared to the IBSC configuration in the radiative limit, from 46.7% to 48.5% with $\Delta E = 260\ meV$. This improvement is gradually decreased when Eg₁ increases, and becomes negligible for $Eg_1 > 2.5\ eV$. We also observe that for $Eg_1 < 1.1\ eV$, without a ratchet, no performance enhancement can be expected by insertion of an IB compared to the single junction. By contrast, the RBSC always exhibits an improvement in our range of study.

The impact of non-idealities, found in nanostructures, is then investigated. First, we study the effect of an absorption width for Eg₃ reduced to $\delta E = 250\ meV$. Such broadenings can be obtained by the formation of minibands in coupled nanostructures. In QWs, widths of about 200 meV have been observed for intersubband transitions [19]. For both the IBSC and the RBSC, an efficiency decrease is observed. This decrease being more pronounced for large Eg₁, and negligible for $Eg_1 < 1.2\ eV$. In the RBSC case, the optimum is found for a ratchet energy $\Delta E = 380\ meV$. Second, a radiative efficiency $r_{rad} = 10^{-3}$ is assumed in addition to the narrow absorption. Such a value can be considered as an upper limit for intersubband transitions at energies larger than the optical phonon energy in QWs [20]. On the one hand, an efficiency reduction is observed for the IBSC, so large that it does not provide any benefit when compared to the best single junction performance. On the other hand, the efficiency in the ratchet configuration is only slightly affected over the whole range of study, the optimum performance being obtained for a ratchet energy increased to $\Delta E = 540\ meV$. This observation indicates that an RBSC, in this configuration, is relatively insensitive to non-radiative recombinations. Because such detrimental recombinations are important in nanostructured IBSC, this constitutes a significant result.

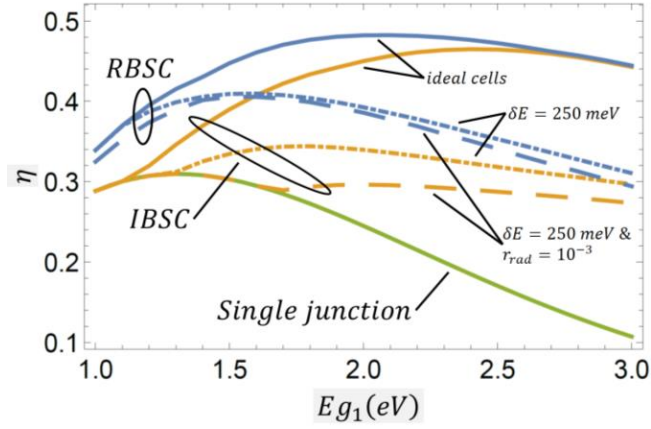

Figure 2 Efficiency as a function of Eg₁ for the single junction (green line), IBSC (yellow lines), and RBSC (blue lines). For the IBSC and the RBSC, the full lines correspond to ideal absorptions and recombinations in the radiative limit. The doted lines correspond to an absorption limited to $\delta E = 250\ meV$, and recombinations in the radiative limit for Eg₃. The dashed lines correspond to $\delta E = 250\ meV$, and a radiative recombination rate of $10^{-3}$ for Eg₃.

For more insight into this remarkable result, we study the interplay of reduced absorption width and radiative recombination ratio in Figure 3(a) and (b). In Figure 3(a), the optimum efficiency is displayed as a function of the absorption width, in the IBSC and RBSC configurations, in the radiative limit and for $r_{rad} = 10^{-3}$. In the radiative limit, we observe that the IBSC efficiency remains constant until the absorption width is reduced to 550 meV, before decreasing linearly until $\delta E = 150\ meV$. Below this value, the implementation of an intermediate level is only detrimental compared to the single junction. No effect is observed for an absorption width above 550 meV, since in that range the optimal bandgap arrangement is obtained for $Eg_2 - Eg_3 < 550\ meV$. Remaining in the radiative limit, we observe that the RBSC is more tolerant to the absorption narrowing than the IBSC, the performance being constant until $\delta E = 450\ meV$ before a linear decrease. Nevertheless, we note that an improvement relative to the SQ limit is possible as soon as the absorption width is non-zero.

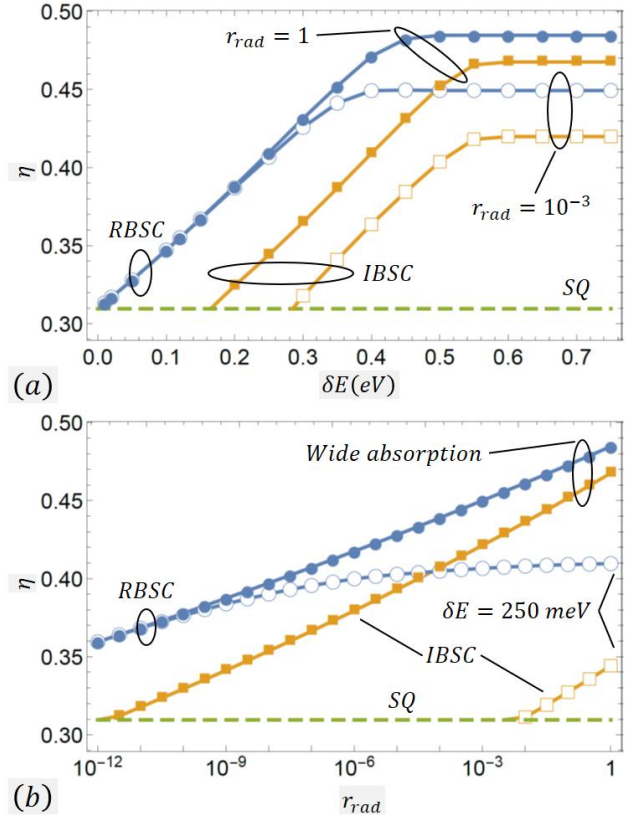

Figure 3 Optimum efficiency of an IBSC (yellow squares) and an RBSC (blue circles): (a) as a function of the absorption width, assuming recombinations at the radiative limit (closed symbols) and a radiative recombination ratio of $10^{-3}$ (open symbols). (b) As a function of the radiative recombination rate, assuming an ideal absorption (closed symbols) and an absorption width reduced to 250 meV (open symbols). In both (a) and (b) the green dashed line represents the Shockley-Queisser limit.

When combining the absorption width narrowing to a non-radiative recombination ratio of $10^{-3}$, we observe that the efficiency of the IBSC is reduced for any absorption width. Below $\delta E = 250\ meV$, no improvement compared to the single junction is expected. On the opposite, the RBSC performance is only impacted for an ideal absorption, while the variation is negligible below $\delta E = 300\ meV$. As a consequence, the efficiency gain thanks to the ratchet is even more apparent, growing from 2.9% absolute for a wide absorption up to 10.8% for $\delta E = 300\ meV$.

In Figure 3(b), the optimum efficiency is displayed as a function of the radiative recombination ratio, in the IBSC and RBSC configurations, for an ideal absorption and for $\delta E = 250\ meV$. For an ideal absorption, the IBSC efficiency decreases monotonously until $r_{rad} = 10^{-12}$ where it reaches the SQ limit. The RBSC is more tolerant to the radiative efficiency deterioration, so that the efficiency gain against the IBSC grows from 1.8% absolute at $r_{rad} = 1$ up to 5% at $r_{rad} = 10^{-12}$. When combining the $r_{rad}$ reduction to an absorption range limited to 250 meV, the IBSC efficiency is strongly reduced: it amounts to 34.4% in the radiative limit, and falls below the SQ limit as soon $r_{rad}$ is lower than $10^{-2}$. By contrast, the ratchet structure drastically reduces the cell sensitivity to the non-radiative recombinations, the efficiency only suffering from a 1% reduction between $r_{rad} = 1$ and $r_{rad} = 10^{-6}$, from 41.0% to 40.0%.

In light of those simulation results, we are now able to draw guidelines for the realization of efficient IBSC with nanostructures, with respect to three parameters: the radiative efficiency and the absorption width of the inter-subband transition, and the possibility of using an electronic ratchet.

On Figure 2 and Figure 3, it was highlighted that below some critical values of $r_{rad}$ and $\delta E$, an IBSC cannot overcome the Shockley-Queisser limit. This observation allows us to define minimum required ($r_{rad}$, $\delta E$), which we plot in Figure 4. It is worth noting that any ($r_{rad}$, $\delta E$) couple, for an RBSC, in the present range, allows expecting efficiencies higher than the single junction limit. As a consequence, this plot equivalently represents the region in which a ratchet mechanism is required for an IBSC to overcome the SQ limit. Nanostructures cells being non-ideal, it is likely that they will stand in that region. As an example, it was calculated in [7] that for InAs QDs in a GaAs matrix, for 5 nm dot-to-dot spacing, the IB to CB absorption width was 14 meV, and the non-radiative recombination ratio would be around $10^{-5}$. Although it is agreed that this system is not ideal, the plot in Figure 4 highlights that it is far from overcoming the SQ limit, without a ratchet mechanism.

We also observe that even in the radiative limit, an absorption width of at least 170 meV is necessary to overcome the SQ limit. This observation leads us to the conclusion that some means to widen the absorption width are mandatory if a ratchet mechanism is not included. This may be obtained with various nanostructure dimensions in a single device, or minibands.

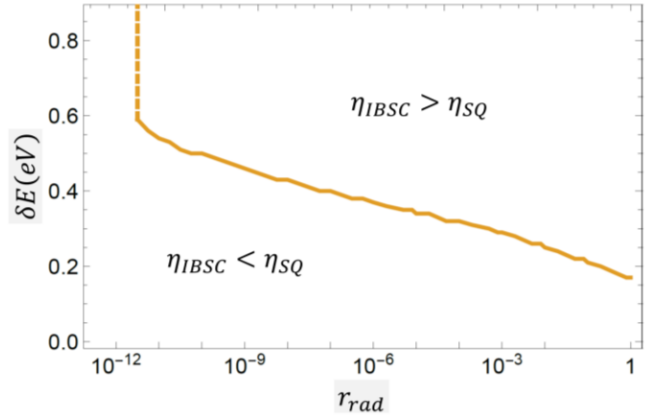

Figure 4 Plot representing the minimum required ($r_{rad}$, $\delta E$) combinations to reach an efficiency larger than the SQ limit, for an IBSC. Within the graph range, any couple can be considered to overcome the SQ limit for an RBSC.

A remarkable feature of an RBSC is its relative insensitivity on the reduction of $r_{rad}$, for narrow absorption, as presented on Figure 3(b). This consideration will have implication regarding the practical implementation of nanostructured IBSC. Indeed, QD have been mainly considered, arguably because they present a true zero density of states in the energy range between the IB and the continuum. Hence, they are expected to exhibit larger radiative recombination ratio compared to QW, in which a continuum of states is available, allowing efficient relaxation via phonon scattering. Nevertheless, QD are not free from non-radiative recombination as well, via intrinsic mechanisms (Auger [7]), or defects [21]. Since both QD and QW will exhibit relatively narrow absorption widths, the radiative recombination ratio benefit of the former on the latter is no more critical in an RBSC. However, a benefit could be obtained by enlarging the absorption width, which can be obtained thanks to a miniband. The growth of packed nanostructures, necessary to build a miniband, seems more favorable in the QW case.

Following this observation, we suggest a structure that creates a ratchet on the conduction band with quantum wells, as represented in Figure 5. This structure uses two sections. The first section consists in large wells, that allow two confined levels, namely IB and CB1. Several coupled wells are necessary to create a miniband that enlarge the absorption width. The second section is made of thinner wells, allowing a single state CB2, that will be engineered to lie at an energy $\Delta E$ below CB1. In order to minimize the recombination from CB2 to IB, an intermediate section is included that will prevent the CB2 wavefunction to extend in section 1. Similarly to what is used in quantum cascade systems, the energy difference between the states connecting CB1 and CB2 should equal the phonon energy to allow efficient relaxation. For practical realization, material combinations exhibiting together large band offsets on the conduction band and reduced offsets on the valence band are required. As an example, InGaAs/AlAsSb lattice matched or strained on InP [22], or InAs/AlAsSb [23] could be used. Other materials have been studied for intersubband transition in the conduction band, among III-V (InGaAs/InAlAs strain balanced on InP [24]), nitrides (GaN/AlN [25]), or II-VI (ZnSe/BeTe [26], (ZnSe/CdSe)/MgSe [27]). Nevertheless, it should be confirmed that the hole transport is still possible in such structures.

Several other ways to implement a ratchet have been suggested in previous literature. Using a quantum cascade has been considered, for a ratchet on the intermediate band [28]. Other strategies include using direct/indirect transitions [14], ferromagnetic compounds [16], 2D materials [29] or organic materials [30]. We also note that previous studies have considered the ratchet mechanism for the up-conversion of low energy photons [31]–[34].

It is also worth noting that other requirements for efficient IBSC, as e.g. the necessity of a partially filled IB, are also required and were not considered in this paper [35]–[37].

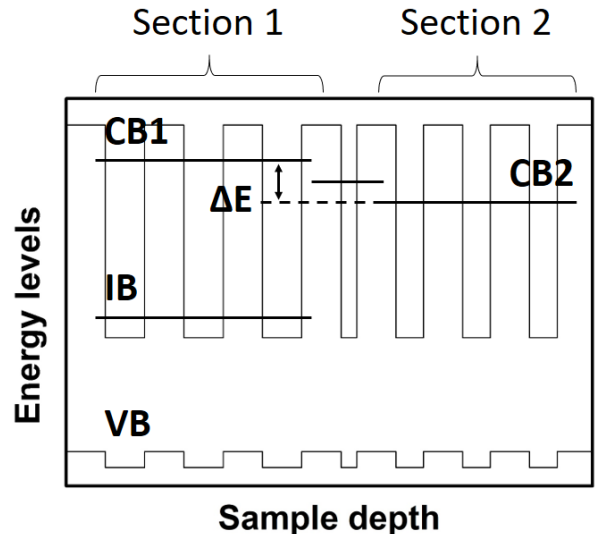

Figure 5 Generic quantum well structure to realize the ratchet mechanism on the conduction band. The left section allows two quantized levels, namely IB and CB1, whereas thinner wells on the right-hand side only allow a single level CB2, at an energy ΔE lower than CB1. In order to prevent the recombination between CB2 and IB, intermediate wells can be inserted.

## IV. CONCLUSION

QDs and QWs have been considered for the realization of IBSC, a concept that has the potential to overcome the Shockley-Queisser limit. Nevertheless, those nanostructures present intrinsic unfavorable properties: intersubband absorptions are only allowed in narrow energy widths, and non-radiative intersubband relaxations are efficient. We have quantified the effect of those two parameters on the cell efficiency, considering in addition the possibility of introducing a ratchet mechanism for mitigating the performance degradation. This allows us to propose guidelines for the design of efficient IBSCs.

We find minimal values of absorption width and non-radiative recombination that are required for efficiencies exceeding the SQ limit when no ratchet is included. Although not theoretically impossible, currently considered nanostructures are not predicted to fulfill those minimal requirements. A key finding is that the introduction of a ratchet mechanism makes the cell highly resilient to the drawbacks of nanostructures, to the point that any combination of absorption width and non-radiative recombination becomes compatible with efficiencies overcoming the Shockley-Queisser limit. A ratchet system therefore appears as a required feature for the practical realization of nanostructured IBSCs.

We observed in addition that the ratchet cell becomes much less sensitive to non-radiative recombination when the absorption width is reduced, which is a remarkable behavior for a solar cell. As a consequence, despite their apparent radiative recombination ratio deficit compared to QDs, it turns out that QWs are good candidates for the realization of efficient IBSC. To take advantage of this observation, a structure realizing a ratchet on the conduction band using QWs is suggested.